\RequirePackage{lineno}

\documentclass[twocolumn,aps,prl,showpacs,superscriptaddress,floatfix]{revtex4}

\usepackage{epsfig}
\usepackage{color}
\newcommand {\Nch}	{N_{\rm ch}}
\newcommand {\Npart}	{N_{\rm part}}

\newcommand {\hic}	{{\sc hic}}
\newcommand {\rhic}	{{\sc rhic}}
\newcommand {\lhc}	{{\sc lhc}}
\newcommand {\dft}	{{\sc dft}}
\newcommand {\ampt}	{{\sc ampt}}
\newcommand {\hijing}	{{\sc hijing}}
\newcommand {\mcg}	{{\sc mcg}}
\newcommand {\ws}	{{\sc ws}}
\newcommand {\cme}	{{\mbox{\sc cme}}}
\newcommand {\cs}	{{\sc cs}}

\newcommand {\RP}	{{\sc rp}}
\newcommand {\PP}	{{\sc pp}}
\newcommand {\EP}	{{\sc ep}}

\newcommand {\Ru}	{$^{96}_{44}$Ru}
\newcommand {\Zr}	{$^{96}_{40}$Zr}
\newcommand {\RuRu}	{RuRu}
\newcommand {\ZrZr}	{ZrZr}
\newcommand {\AuAu}	{AuAu}

\newcommand {\rt}	{r_{\perp}}

\newcommand {\rhows}	{\rho_{_{\rm WS}}}
\newcommand {\psiRP}	{\psi_{_{\rm RP}}}
\newcommand {\psiPP}	{\psi_{_{\rm PP}}}
\newcommand {\psiEP}	{\psi_{_{\rm EP}}}
\newcommand {\psiB}	{\psi_{_{\rm B}}}

\newcommand {\etwo}	{\epsilon_2}
\newcommand {\epsi}	{\etwo\{\psi\}}
\newcommand {\eRP}	{\etwo\{\psiRP\}}
\newcommand {\ePP}	{\etwo\{\psiPP\}}
\newcommand {\vv}	{v_2}
\newcommand {\vRP}	{\vv\{\psiRP\}}
\newcommand {\vEP}	{\vv\{\psiEP\}}
\newcommand {\vEPobs}	{\vv\{\psiEP^{\rm rec}\}}
\newcommand {\psiEPobs}	{\psiEP^{\rm rec}}
\newcommand {\res}	{\mathcal{R}_{\rm EP}}

\newcommand {\rbf}	{{\bf r}}
\newcommand {\Bbf}	{{\bf B}}
\newcommand {\Bsq}	{\overline{B_{\rm sq}}}
\newcommand {\Bpsi}	{\Bsq\{\psi\}}
\newcommand {\BRP}	{\Bsq\{\psiRP\}}
\newcommand {\BPP}	{\Bsq\{\psiPP\}}
\newcommand {\BEP}	{\Bsq\{\psiEP\}}
\newcommand {\BEPobs}	{\Bsq\{\psiEP^{\rm rec}\}}
\newcommand {\Bcos}	{(eB(\rbf,0)/m_\pi^2)^2\cos2(\psiB-\psi)}

\newcommand {\gOS}	{\gamma_{_{\rm OS}}}
\newcommand {\gSS}	{\gamma_{_{\rm SS}}}
\newcommand {\dg}	{\Delta\gamma}

\newcommand {\mean}[1]	{\langle #1\rangle}

\begin{document}
\title{Importance of isobar density distributions on the chiral magnetic effect search}
\author{Hao-jie Xu}
\affiliation{School of Science, Huzhou University, Huzhou, Zhejiang 313000, China}
\author{Xiaobao Wang}
\affiliation{School of Science, Huzhou University, Huzhou, Zhejiang 313000, China}
\author{Hanlin Li}
\affiliation{College of Science, Wuhan University of Science and Technology, Wuhan, Hubei 430065, China}
\author{Jie Zhao}
\affiliation{Department of Physics and Astronomy, Purdue University, West Lafayette, Indiana 47907, USA}
\author{Zi-Wei Lin}
\affiliation{Department of Physics, East Carolina University, Greenville, North Carolina 27858, USA}
\affiliation{Key Laboratory of Quarks and Lepton Physics (MOE) and Institute of Particle Physics, Central China Normal University, Wuhan, Hubei 430079, China}
\author{Caiwan Shen}
\affiliation{School of Science, Huzhou University, Huzhou, Zhejiang 313000, China}
\author{Fuqiang Wang}
\email{fqwang@zjhu.edu.cn}
\affiliation{School of Science, Huzhou University, Huzhou, Zhejiang 313000, China}
\affiliation{Department of Physics and Astronomy, Purdue University, West Lafayette, Indiana 47907, USA}

\date{\today}

\begin{abstract}
Under the approximate chiral symmetry restoration, quark interactions with topological gluon fields  in quantum chromodynamics can induce chirality imbalance and parity violation in local domains. An electric charge separation (\cs) could be generated along the direction of a strong magnetic field ($\Bbf$), a phenomenon called the chiral magnetic effect (\cme). \cs\ measurements by azimuthal correlators 
are contaminated by major backgrounds from elliptic flow anisotropy ($\vv$). Isobaric \Ru+\Ru\ and \Zr+\Zr\ collisions have been proposed to identify the \cme\ (expected to differ between the two systems) out of the backgrounds (to be almost the same). We show, by using the density functional theory calculations of the proton and neutron distributions, 
that these expectations 
may not hold as originally anticipated because the two systems may have sizable differences in eccentricity and $\vv$.
\end{abstract}


\maketitle

{\em Introduction.}
Due to vacuum fluctuations, topological gluon fields can emerge in quantum chromodynamics (QCD)~\cite{Lee:1974ma}. 
The interactions of quarks with those gluon fields can induce chirality imbalance and parity violation in local domains
under the approximate chiral symmetry restoration~\cite{Lee:1974ma,Morley:1983wr,Kharzeev:1998kz,Kharzeev:2007jp}, 
likely achieved in relativistic heavy ion collisions (\hic) at BNL's Relativistic Heavy Ion Collider (\rhic)~\cite{Arsene:2004fa,Back:2004je,Adams:2005dq,Adcox:2004mh} and CERN's Large Hadron Collider (\lhc)~\cite{Muller:2012zq}. A chirality imbalance could lead to an electric current, or charge separation (\cs) 
in the direction of a strong magnetic field ($\Bbf$)~\cite{Kharzeev:2007jp}. 
This phenomenon is called the chiral magnetic effect (\cme)~\cite{Fukushima:2008xe}. 
Searching for the \cme\ is one of the most active research in \hic~\cite{Abelev:2009ac,Abelev:2009ad,Abelev:2012pa,Adamczyk:2013hsi,Adamczyk:2014mzf,Adamczyk:2013kcb,Khachatryan:2016got,Sirunyan:2017quh,Acharya:2017fau}. 
The \cme\ is not specific to QCD but a macroscopic phenomenon generally arising from quantum anomalies~\cite{Kharzeev:2015znc}. It is a subject of interest for a wide range of physics communities; such phenomena have been observed in magnetized relativistic matter in three-dimensional Dirac and Weyl materials~\cite{Li:2014bha,Lv:2015pya,Huang:2015eia}. 

In \hic\ the \cs\ is commonly measured by the three-point correlator~\cite{Voloshin:2004vk}, $\gamma\equiv\cos(\phi_\alpha+\phi_\beta-2\psiRP)$, where $\phi_\alpha$ and $\phi_\beta$ are the azimuthal angles of two charged particles, and $\psiRP$ is that of the reaction plane (\RP, spanned by the impact parameter and beam directions) to which the $\Bbf$ produced by the incoming protons is perpendicular on average~\cite{Kharzeev:2004ey,Bzdak:2011yy,Deng:2012pc,Bloczynski:2012en}. 
Often a third particle azimuthal angle is used in place of $\psiRP$ with a resolution correction~\cite{Abelev:2009ac,Abelev:2009ad}.
Because of charge-independent backgrounds, such as correlations from global momentum conservation, 
the correlator difference between opposite-sign ({\sc os}) and same-sign ({\sc ss}) pairs, $\dg\equiv\gOS-\gSS$, is used. 
Positive $\dg$ signals, consistent with the \cme-induced \cs\ perpendicular to the \RP, have been observed~\cite{Abelev:2009ac,Abelev:2009ad,Adamczyk:2014mzf,Adamczyk:2013hsi,Abelev:2012pa}. 
The signals are, however, inconclusive because of a large charge-dependent background arising from particle correlations (e.g.~resonance decays) coupled with the elliptic flow anisotropy ($\vv$)~\cite{Wang:2009kd,Bzdak:2009fc,Schlichting:2010qia}. 
Take $\rho^0\rightarrow\pi^+\pi^-$ as an example~\cite{Voloshin:2004vk,Wang:2016iov}. Because of the $\vv$ of $\rho$, more {\sc os} pairs align in the \RP\ than $\Bbf$ direction, leading to a sizable signal: 
$\dg\propto\mean{\cos(\alpha+\beta-2\phi_{\rho})\cos2(\phi_{\rho}-\psiRP)}\propto v_{2,\rho}$
~\cite{Wang:2016iov}. In other words, the $\gOS$ variable is ambiguous between a \cme-induced back-to-back pair (\cs) perpendicular to the \RP\ and a resonance-decay pair (charge alignment) along the \RP~\cite{Wang:2009kd,Bzdak:2009fc,Adamczyk:2013kcb}.

There have been many attempts to reduce/eliminate the $\vv$-induced backgrounds~\cite{Ajitanand:2010rc,Bzdak:2011np,Adamczyk:2013kcb,Wang:2016iov,Zhao:2017nfq}. 
STAR~\cite{Adamczyk:2013kcb} 
found a charge asymmetry signal to linearly depend on the event-by-event $\vv$ of final-state particles, suggesting a background dominance. ALICE~\cite{Acharya:2017fau} and CMS~\cite{Sirunyan:2017quh} divided their data from each collision centrality according to their event-by-event $\vv$, and found the $\dg$ signal to be proportional to $\vv$, consistent with a null \cme. 

To better control the background, isobaric collisions of \Ru+\Ru\ (\RuRu) and \Zr+\Zr\ (\ZrZr) have been proposed~\cite{Voloshin:2010ut}. One expects their backgrounds to be almost equal because of the same mass number, while the atomic numbers, hence $\Bbf$, differ by 10\%. 
This is verified by {\em Monte Carlo} Glauber (\mcg) calculations~\cite{Deng:2016knn} using the Woods-Saxon (\ws) density profile,
\begin{equation}
\rhows(r,\theta)\propto\left(1+\exp[(r-R_0[1+\beta_2Y_2^0(\theta)])/a]\right)^{-1}\;,
\label{eq:ws}
\end{equation}
where 
$R_0=5.085$~fm and 5.020~fm are used for Ru and Zr, respectively, $a=0.46$~fm, and $Y_2^0$ is a spherical harmonic. 
The deformity quadrupole parameter $\beta_2$ has large uncertainties; current knowledge suggests two contradicting sets of values~\cite{Deng:2016knn}, 0.158 (Ru) and 0.080 (Zr)~\cite{Raman:1201zz,Pritychenko:2013gwa} vis a vis 0.053 (Ru) and 0.217 (Zr)~\cite{Moller:1993ed,Kumar:2014ypa,Moller:2015fba}. This would yield a less than $\pm2$\% difference in eccentricity ($\etwo$), hence a residual $\vv$ background, between \RuRu\ and \ZrZr\ collisions in the 20-60\% centrality range~\cite{Deng:2016knn}.
$\Bbf^2$, to which the \cme\ strength in $\dg$ is proportional, differs by approximately 15\% (not the simple 19\% because of the slightly smaller $R_0$ value used for Zr than Ru)~\cite{Deng:2016knn}.
As a net result, the \cme\ signal to background ratio would be improved by over a factor of seven in comparative measurements between \RuRu\ and \ZrZr\ collisions than in each of them individually~\cite{Deng:2016knn}.
The isobaric collisions are planned for 2018 at \rhic; they would yield a \cme\ signal of $5\sigma$ significance with the projected data volume, 
if one assumes that the \cme\ contributes 1/3 of the current $\dg$ measurement in \AuAu\ collisions~\cite{Deng:2016knn}.

However, there can be non-negligible deviations of the Ru and Zr nuclear densities from \ws. The purpose of this Letter is to investigate those deviations and their effects on the sensitivity of isobaric collisions for the \cme\ search. 

\begin{figure}[hbt]
  \begin{center}
    \includegraphics[width=0.3\textwidth]{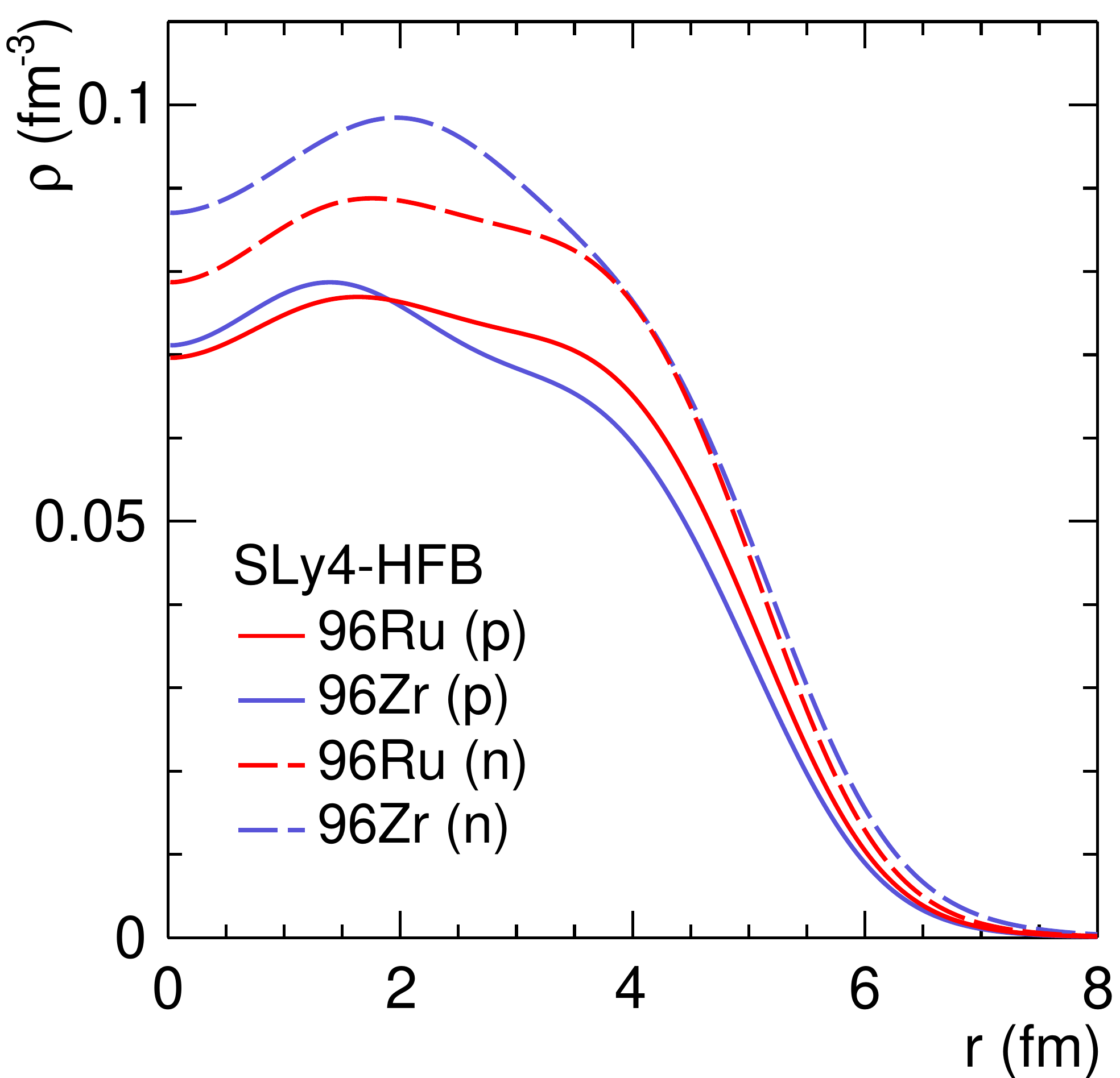}
  \end{center}
  \vspace{-0.2in}
  \caption{(Color online) Proton and neutron density distributions of the \Ru\ and \Zr\ nuclei, assumed spherical, calculated by the \dft\ method.}
  \label{fig:rho}
\end{figure}

{\em Nuclear densities.}
Because of the different numbers of protons--which suffer from Coulomb repulsion--and neutrons, the structures of the \Ru\ and \Zr\ nuclei must not be identical. Measurements of their charge and mass densities are, however, scarce~\cite{Raman:1201zz,Pritychenko:2013gwa,Deng:2016knn}. Their knowledge requires theoretical calculations~\cite{Moller:1993ed,Kumar:2014ypa,Moller:2015fba,Wang:2016rqh}. 
Much of the theoretical understanding of proton and neutron distributions in nuclei came, so far, from density functional theory (\dft)~\cite{Bender:2003jk,Erler:2012xxx}. 
While {\em ab initio} methods have been employed to calculate nuclear structures up to $^{48}$Ca~\cite{Hagen:2015yea,Ruiz:2016gne}, 
\dft\ is at present the only microscopic approach which can be applied throughout the entire nuclear chart~\cite{Kortelainen:2014uma}. 
It employs energy density functionals which incorporate complex many-body correlations into functionals that are primarily constrained by global nuclear properties such as binding energies and radii~\cite{Bender:2003jk,Erler:2012xxx,Hagen:2015yea}.
By using \dft, we calculate the Ru and Zr proton and neutron distributions using the well-known SLy4 mean field~\cite{Chabanat:1997un} including pairing correlations (Hartree-Fock-Bogoliubov, HFB approach)~\cite{Dreizler1990nuclear,ring2000nuclear,Bender:2003jk,Wang:2016rqh}. 
The calculated ground-state proton and nucleon (proton+neutron) densities, assumed spherical, are shown in Fig.~\ref{fig:rho}.
Protons in Zr are more concentrated in the core, while protons in Ru, 10\% more than in Zr, are pushed more toward outer regions. The neutrons in Zr, four more than in Ru, are more concentrated in the core but also more populated on the nuclear skin. 

Theoretical uncertainties are estimated by using different sets of density functionals, SLy5~\cite{Chabanat:1997un} and SkM*~\cite{Bartel:1982ed} for the mean field, with and without pairing (HFB/HF)~\cite{Dreizler1990nuclear,ring2000nuclear,Bender:2003jk}, and found to be small. 
The deformities of Ru and Zr are uncertain, allowed by a wide range of possibilities~\cite{Deng:2016knn,Raman:1201zz,Pritychenko:2013gwa,Moller:1993ed,Kumar:2014ypa,Moller:2015fba}. Our \dft\ calculations indicate that their ground states are soft against deformation and can be nearly spherical. Their densities are calculated with the allowed extreme values of $\beta_2$ (0.158 for Ru and 0.217 for Zr~\cite{Raman:1201zz,Pritychenko:2013gwa,Moller:1993ed,Kumar:2014ypa,Moller:2015fba}). They yield the largest uncertainties on our results. 

\begin{figure}[hbt]
  \begin{center}
    \includegraphics[width=0.235\textwidth]{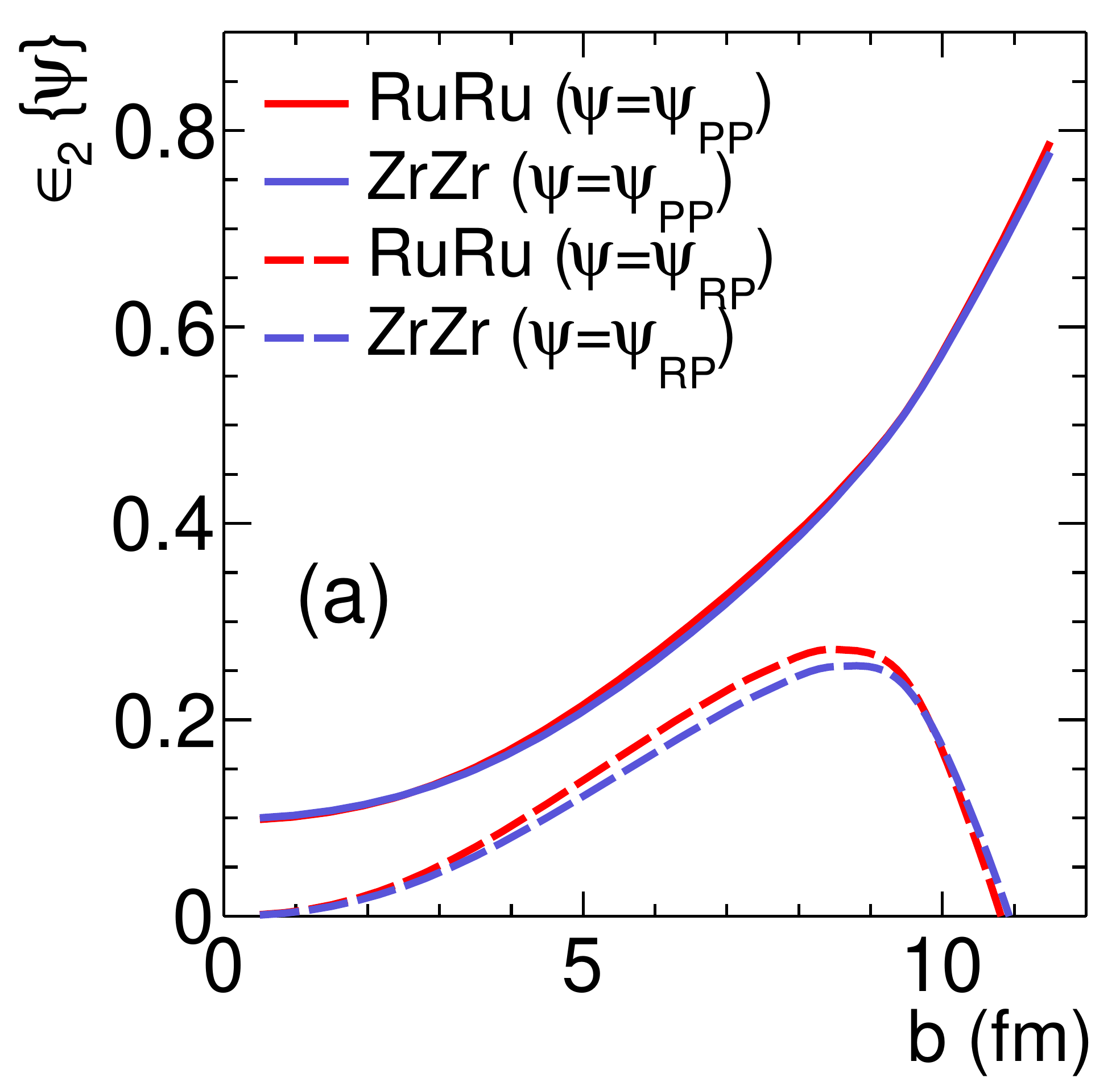}
    \includegraphics[width=0.235\textwidth]{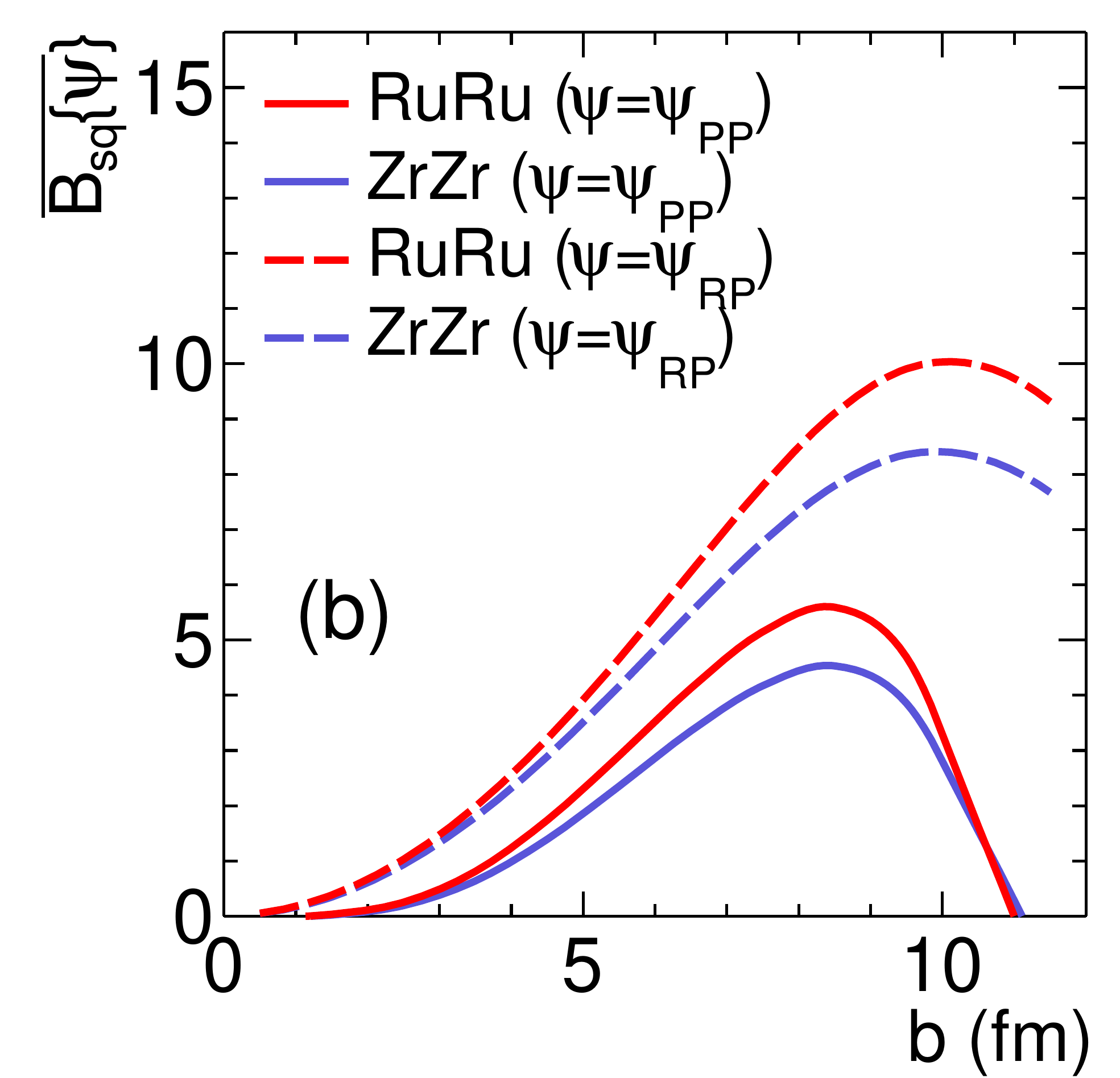}
  \end{center}
  \vspace{-0.2in}
  \caption{(Color online) (a) $\epsi$ and (b) $\Bpsi$ with respect to $\psi=\psiRP$ and $\psiPP$ as functions of $b$ in \RuRu\ and \ZrZr\ collisions, calculated by \mcg\ with the \dft\ densities in Fig.~\ref{fig:rho}.}
  \label{fig:Glauber}
\end{figure}

{\em Eccentricity and magnetic field.}
The $\etwo$ of the transverse overlap geometry in \RuRu\ and \ZrZr\ collisions is calculated event-by-event with \mcg~\cite{Alver:2006wh,Miller:2007ri,Rybczynski:2011wv,Xu:2014ada,Zhu:2016puf}, using the nucleon densities in Fig.~\ref{fig:rho}, by
\begin{equation}
  \ePP_{\rm evt}e^{i2\psiPP}=\mean{\rt^2 e^{i2\phi_{r}}}/\mean{\rt^2}\;.
  \label{eq:ePP}
\end{equation}
Here $\mean{...}$ denotes the per-event average; ($\rt$,$\phi_{r}$) is the polar coordinate of each initial participant nucleon in the transverse plane, whose origin $\rbf=0$ is taken to be the center of mass of all participant nucleons. 
The $\etwo$ is the average over many events, $\ePP\equiv\mean{\ePP_{\rm evt}}$. 
The nucleon-nucleon cross-section is taken to be 42~mb~\cite{Abelev:2008ab,Xu:2014ada} with the ``Gaussian'' approach~\cite{Rybczynski:2011wv}; a minimum nucleon-nucleon separation of 0.4~fm is required~\cite{Abelev:2008ab,Xu:2014ada}; uncertainties on these values have negligible effect on our results.
The $\ePP$ is the eccentricity with respect to the participant plane (\PP). Due to finite number effect, the \PP\ azimuthal angle $\psiPP$ fluctuates about the \RP\ azimuthal angle, $\psiRP$ (fixed at 0)~\cite{Alver:2006wh}; the $\etwo$ of the averaged overlap geometry is
\begin{equation}
  \eRP=\mean{\ePP\cos2(\psiPP-\psiRP)}\;.
  \label{eq:eRP}
\end{equation}
The $\ePP$ and $\eRP$ calculated using the \dft\ densities are shown in Fig.~\ref{fig:Glauber}(a) as functions of the impact parameter ($b$). 

$\Bbf(\rbf,t=0)$ is calculated 
for \RuRu\ and \ZrZr\ collisions using the proton densities in Fig.~\ref{fig:rho}. The calculations follow Ref.~\cite{Deng:2012pc,Deng:2014uja}, with a finite proton radius (0.88~fm~\cite{Deng:2014uja} is used but the numeric value is not critical) to avoid the singularity at zero relative distance.  
The relevant quantity~\cite{Deng:2016knn} for the \cme\ strength in a $\dg$ measurement, with respect to an azimuth $\psi$, is the event average, $\Bpsi\equiv\mean{\Bpsi_{\rm evt}}$;
\begin{eqnarray}
\Bpsi_{\rm evt}&\equiv&\int\Npart^2(\rbf)\Bcos d\rbf\nonumber\\
&&\left/\int\Npart^2(\rbf)d\rbf\right.\;,
\label{eq:Bsq}
\end{eqnarray}
where $\Npart(\rbf)$ is the transverse density of participant nucleons. The average is weighted by $\Npart^2$ because $\dg$ is a pair-wise observable; our results are, however, only weakly sensitive to the $\Npart$-weighting power.
Figure~\ref{fig:Glauber}(b) shows $\BRP$ and $\BPP$ calculated using the \dft\ densities.
Since $\Bbf$ in non-central \hic\ comes primarily from the spectator protons, its event-averaged direction is perpendicular to $\psiRP$, not $\psiPP$. $\BPP$ is a projection of and hence always smaller than $\BRP$, in contrast to the case for $\etwo$ in Eq.~(\ref{eq:eRP}).

\begin{figure}[hbt]
  \begin{center}
    \hspace*{-0.02\textwidth}
    \includegraphics[width=0.5\textwidth]{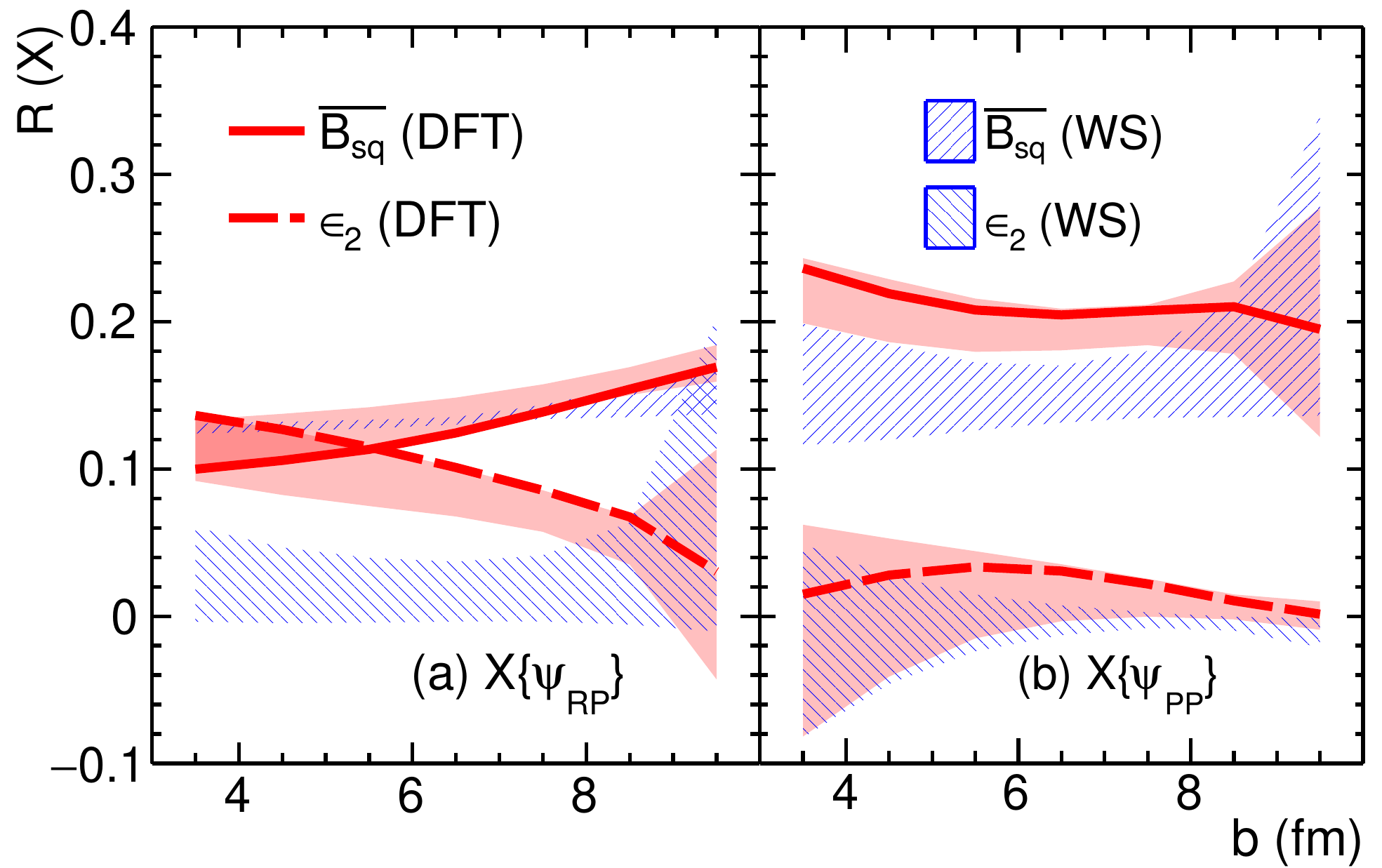}
  \end{center}
  \vspace{-0.2in}
  \caption{(Color online) Relative differences between \RuRu\ and \ZrZr\ collisions in $\epsi$ and $\Bpsi$ with respect to (a) $\psi=\psiRP$ and (b) $\psi=\psiPP$, using the \dft\ densities in Fig.~\ref{fig:rho}. The shaded areas correspond to \dft\ density uncertainties from Ru and Zr deformities; the hatched areas show the corresponding results using \ws\ of Eq.~(\ref{eq:ws}).}
  \label{fig:R}
\end{figure}

For the \cme\ search with isobaric collisions, the relative differences in $\etwo$ and $\Bsq$ are of importance. Figure~\ref{fig:R} shows the relative differences $R(\ePP)$, $R(\eRP)$, $R(\BPP)$, and $R(\BRP)$; $R(X)$ is defined as~\cite{Deng:2016knn}
\begin{equation}
R(X)\equiv2(X_{\rm\RuRu}-X_{\rm\ZrZr})/(X_{\rm\RuRu}+X_{\rm\ZrZr})\;
\end{equation}
where $X_{\rm\RuRu}$ and $X_{\rm\ZrZr}$ are the $X$ values in \RuRu\ and \ZrZr\ collisions, respectively. The thick solid curves are the default results with the \dft\ densities in Fig.~\ref{fig:rho}. 
The shaded areas correspond to theoretical uncertainties bracketed by the two \dft\ density cases where Ru is deformed with $\beta_2=0.158$ and Zr is spherical and where Ru is spherical and Zr is deformed with $\beta_2=0.217$. 
The hatched areas represent our results using \ws\ densities in Eq.~(\ref{eq:ws}) with the above two cases of nuclear deformities.

{\em Event plane and elliptic anisotropy.}
We investigate whether our density profiles would, in a dynamical model, lead to a final-state $\vv$ difference between \RuRu\ and \ZrZr\ collisions and whether the $\Bsq$ difference preserves with respect to the event plane (\EP) reconstructed from the final-state particle momenta. 
We employ A Multi-Phase Transport (\ampt) model with ``string melting''~\cite{Lin:2001zk,Lin:2004en}, 
which can reasonably reproduce heavy ion bulk data at \rhic\ and the \lhc~\cite{Lin:2014tya,Ma:2016fve}. 
The initial condition of \ampt\ is taken from \hijing~\cite{Wang:1991hta}. 
We implement our \dft\ nuclear densities into the \hijing\ component in \ampt. 
The string-melting \ampt\ converts the \hijing-produced initial hadrons into their valence quarks~\cite{Lin:2001zk,Lin:2004en}, 
which further evolve via two-body elastic scatterings~\cite{Zhang:1997ej}. 
The Debye-screened differential cross-section $d\sigma/dt\propto\alpha_s^2/(t-\mu_D^2)^2$~\cite{Lin:2004en} is used, with strong coupling constant $\alpha_s=0.33$ and screening mass $\mu_D=2.265$/fm (so the total cross section is $\sigma=3$~mb). 
After quarks stop interacting, a simple coalescence model is applied to describe the hadronization process that converts quarks into hadrons ~\cite{Lin:2004en}. 
We switch off subsequent hadronic scatterings in \ampt, as was done in Ref.~\cite{Ma:2011uma,Zhao:2017nfq}; 
while responsible for the majority of the $\vv$ mass splitting, they are not important for the $\vv$ magnitude~\cite{Li:2016flp,Li:2016ubw}. 

The \ampt\ version and parameter values used in the present work are the same as those used earlier for \rhic\ collisions in~\cite{Lin:2014tya,Ma:2016fve,He:2015hfa,Li:2016flp,Li:2016ubw}. A total on the order of 50 million minimum-bias events each are simulated for \RuRu\ and \ZrZr\ collisions with $b$ from 0 to 12~fm. 
The charged particle (hereafter referring to $\pi^\pm$, $K^\pm$, $p$, and $\bar{p}$ within pseudorapidity $|\eta|<1$) multiplicity ($\Nch$) distribution in \RuRu\ has a slightly higher tail than that in \ZrZr. The difference is insignificant; for example, the 20-60\% centrality corresponds to the $\Nch$ range of 62-273 and 61-271 in \RuRu\ and \ZrZr, respectively.

\begin{figure}[hbt]
  \begin{center}
    \includegraphics[width=0.35\textwidth]{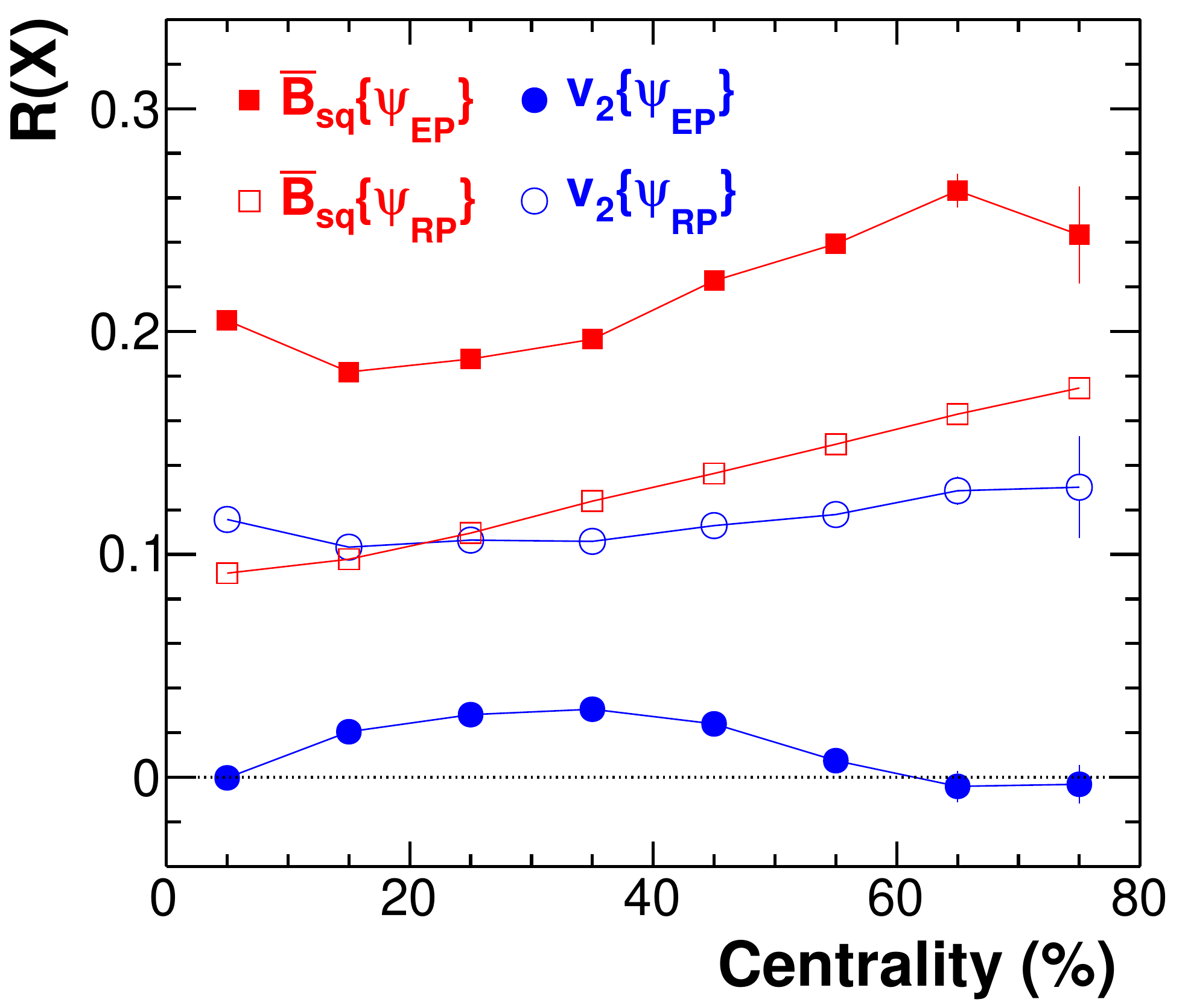}
  \end{center}
  \vspace{-0.2in}
  \caption{(Color online) Relative differences between \RuRu\ and \ZrZr\ collisions as functions of centrality in $\vv\{\psi\}$ (of charged particles in $|\eta|<1$) and $\Bpsi$ with respect to $\psi=\psiRP$ and $\psiEP$, simulated by \ampt\ with the \dft\ densities in Fig.~\ref{fig:rho}. 
}
  \label{fig:AMPT}
\end{figure}

The \EP\ azimuthal angle is reconstructed similar to Eq.~(\ref{eq:ePP}), 
$\vEPobs_{\rm evt}e^{i2\psiEPobs}=\mean{e^{i2\phi}}$,
but with final-state charged particle azimuthal angle $\phi$ in momentum space. 
The $\vv$
is corrected by the \EP\ resolution ($\res$), $\vEP=\mean{\vEPobs_{\rm evt}}/\res$~\cite{Poskanzer:1998yz}.
The $\vv$ with respect to the \RP\ is simply given by
$\vRP=\mean{\cos2(\phi-\psiRP)}$, where $\psiRP=0$ is fixed.
The $\vRP$ and $\vEP$ are found to follow the $b$-dependence of the eccentricities calculated in \ampt\ (which are consistent with those from our \mcg). 
$\Bbf(\rbf,t=0)$ 
is also computed from the initial incoming protons in \ampt, as done in \mcg, for $\BRP$ and $\BEP\equiv\BEPobs/\res$. $\BRP$ is consistent with that calculated by \mcg; $\BEP$ is found to be similar to $\BPP$.
Figure~\ref{fig:AMPT} shows $R(\vRP), R(\vEP), R(\BRP)$, and $R(\BEP)$ from \ampt\ as functions of centrality, determined from the $\Nch$ distributions. The general trends are similar to those in Fig.~\ref{fig:R}.

{\em Discussions and Summary.}
Isobaric \RuRu\ and \ZrZr\ collisions were proposed to help search for the \cme\ for their expected different $\Bbf$ and equal $\etwo$~\cite{Voloshin:2010ut}. These expectations are qualitatively verified by \mcg\ calculations using \ws\ in Eq.~(\ref{eq:ws})~\cite{Deng:2016knn}. 
We have generally reproduced those results with our \mcg, which are shown as the hatched areas in Fig.~\ref{fig:R}. 
Our $\Bsq$ is an average over the transverse overlap area while in Ref.~\cite{Deng:2016knn} it is that at $({\bf 0},0)$. The \RuRu-\ZrZr\ differences in these two $\Bsq$ quantities are less similar for our \dft\ calculated densities than for the more regular \ws.

$R(\BRP)$ is slightly smaller for the \dft\ density profiles than for \ws\ at small $b$. 
This is consistent with the hierarchy in the charge radii differences between Ru and Zr: $\sqrt{\mean{r^2}}=4.327$~fm and 4.271~fm from \dft, and those from Eq.~(\ref{eq:ws}).
The Zr mass radius (4.366~fm from \dft) is, on the other hand, not smaller than Ru's (4.343~fm), making the $\Nch$ distribution tail in \RuRu\ slightly higher than in \ZrZr, opposite to Ref.~\cite{Deng:2016knn}.
For \cs\ measurements with respect to the 2nd order harmonic \EP, however, it is the $\BPP$, not the $\BRP$, that matters. 
$R(\BPP)$ from the \dft\ densities is larger than its \ws\ counterpart.
It is interesting to note that $R(\BPP)$ is always larger than $R(\BRP)$; it is found to arise from a better alignment of $\psiPP$ with $\psiRP$ in \RuRu, by about 10\%, than in \ZrZr. This is because the Ru mass density outweights the Zr's in the outer region while Zr is more concentrated at the core, making the $\psiPP$ better determined in \RuRu\ than in \ZrZr. 

The \dft\ calculated densities introduce a large $\eRP$ difference, as large as that in $\BRP$. This means that, with respect to \RP, the premise of isobaric collisions to help identify the \cme\ does not hold. 
The \dft\ calculated densities introduce a sizable $\ePP$ difference, up to $R(\ePP)\approx3.7\%$ at $b\approx5$~fm (Fig.~\ref{fig:R}(b) dashed curves), and an average $\vv$ difference $R(\vEP)\approx2.7\pm0.1\%$ in 20-60\% centrality (Fig.~\ref{fig:AMPT} filled circles). 
Although this $\vv$ difference is significantly smaller than the difference in the magnetic field, it can have a sizable effect on the isobar difference because of the background dominance in the experimental $\dg$ measurement. For example, suppose 10\% of the measured $\dg$ comes from the \cme\ signal, then the $\Bsq$ difference of 20\% would introduce only a 2\% effect while the $\vv$ difference gives a 2.4\% effect. In other words, one could measure a 4.4\% isobar difference in $\dg$, out of which more than half is due to background. 
The sizable $\ePP$ and $\vEP$ difference weakens the power of isobaric collisions to search for the \cme. 
A direct calculation of the $\gamma$ correlators with realistic backgrounds and an assumed CME signal would be valuable to the \cme\ search.
%
%
Experimentally the $\vv$ will be measured, which would gauge what the geometry difference likely to be between \RuRu\ and \ZrZr. 
Our work suggests that a sizable $\vv$ difference up to $\sim3$\% is likely and one needs to carefully examine $\vv$ and $\dg$ measurements in assessing the possible \cme\ signal.

In summary, topological charge fluctuations are a fundamental property of QCD, which could lead to the chiral magnetic effect (\cme) and charge separation (\cs) in relativistic heavy ion collisions. Experimental \cs\ measurements have suffered from major backgrounds from resonance decays coupled with elliptic flow anisotropy ($\vv$). To reduce background effects, isobaric \Ru+\Ru\ and \Zr+\Zr\ collisions have been proposed where the $\vv$-induced backgrounds are expected to be similar while the \cme-induced signals to be different. In this Letter, the proton and neutron density distributions of \Ru\ and \Zr\ are calculated using the energy density functional theory (\dft). They are then implemented in the {\em Monte Carlo} Glauber (\mcg) model to calculate the eccentricities ($\etwo$) and magnetic fields ($\Bbf$); the \dft\ densities are implemented in A Multi-Phase Transport (\ampt) model to simulate the $\vv$. 
It is found that those nuclear densities, together with the Woods-Saxon (\ws) densities, yield wide ranges of differences in $\Bsq$ with respect to the participant plane (\PP) and the reaction plane (\RP). 
It is further found that those nuclear densities introduce, in contrast to \ws, comparable differences in $\eRP$ ($\vRP$) and $\BRP$ with respect to the reaction plane (\RP), diminishing the premise of isobaric collisions to help identify the \cme. With respect to the participant plane (\PP), the $\ePP$ ($\vEP$) difference can still be sizable, as large as $\sim3$\%, possibly weakening the power of isobaric collisions for the \cme\ search. 

Since the \dft\ calculation of the matter radius is smaller for Ru and Zr, the produced particle multiplicity distribution would have a higher tail in \RuRu\ than in \ZrZr, as predicted by \ampt. This can be checked against results using density distributions of larger \Ru\ than \Zr\ radius, such as \ws\ densities using charge radii in place of matter radii. 
We further predict, using the \dft\ calculated density distributions, that the $\vv$ difference between \RuRu\ and \ZrZr\ with respect to the \RP\ is larger than that with respect to the \PP\ by an absolute 8\%, insensitive to uncertainties in the nuclear deformities, while it is practically zero for \ws. This can be experimentally tested by the upcoming isobaric collisions; a confirmation would be a good indication of the validity of the density distributions calculated here for the Ru and Zr nuclei.
Our study would then be a valuable guidance to the experimental isobaric collision program at \rhic.

{\em Acknowledgments.}
FW thanks B.~Alex Brown for useful discussions.
This work was supported in part by the National Natural Science Foundation of China under Grants No. 11647306, 11747312, U1732138, 11505056 and 11605054, 11628508, and US~Department of Energy Grant No.~DE-SC0012910. 

\bibliography{../../../ref}

\newcommand{\sNN}{$\sqrt{s_{NN}}$}
\begin{thebibliography}{70}
\expandafter\ifx\csname natexlab\endcsname\relax\def\natexlab#1{#1}\fi
\expandafter\ifx\csname bibnamefont\endcsname\relax
  \def\bibnamefont#1{#1}\fi
\expandafter\ifx\csname bibfnamefont\endcsname\relax
  \def\bibfnamefont#1{#1}\fi
\expandafter\ifx\csname citenamefont\endcsname\relax
  \def\citenamefont#1{#1}\fi
\expandafter\ifx\csname url\endcsname\relax
  \def\url#1{\texttt{#1}}\fi
\expandafter\ifx\csname urlprefix\endcsname\relax\def\urlprefix{URL }\fi
\providecommand{\bibinfo}[2]{#2}
\providecommand{\eprint}[2][]{\url{#2}}

\bibitem[{\citenamefont{Lee and Wick}(1974)}]{Lee:1974ma}
\bibinfo{author}{\bibfnamefont{T.}~\bibnamefont{Lee}} \bibnamefont{and}
  \bibinfo{author}{\bibfnamefont{G.}~\bibnamefont{Wick}},
  \bibinfo{journal}{Phys.Rev.} \textbf{\bibinfo{volume}{D9}},
  \bibinfo{pages}{2291} (\bibinfo{year}{1974}).

\bibitem[{\citenamefont{Morley and Schmidt}(1985)}]{Morley:1983wr}
\bibinfo{author}{\bibfnamefont{P.~D.} \bibnamefont{Morley}} \bibnamefont{and}
  \bibinfo{author}{\bibfnamefont{I.~A.} \bibnamefont{Schmidt}},
  \bibinfo{journal}{Z. Phys.} \textbf{\bibinfo{volume}{C26}},
  \bibinfo{pages}{627} (\bibinfo{year}{1985}).

\bibitem[{\citenamefont{Kharzeev et~al.}(1998)\citenamefont{Kharzeev, Pisarski,
  and Tytgat}}]{Kharzeev:1998kz}
\bibinfo{author}{\bibfnamefont{D.}~\bibnamefont{Kharzeev}},
  \bibinfo{author}{\bibfnamefont{R.}~\bibnamefont{Pisarski}}, \bibnamefont{and}
  \bibinfo{author}{\bibfnamefont{M.~H.} \bibnamefont{Tytgat}},
  \bibinfo{journal}{Phys.Rev.Lett.} \textbf{\bibinfo{volume}{81}},
  \bibinfo{pages}{512} (\bibinfo{year}{1998}), \eprint{hep-ph/9804221}.

\bibitem[{\citenamefont{Kharzeev et~al.}(2008)\citenamefont{Kharzeev, McLerran,
  and Warringa}}]{Kharzeev:2007jp}
\bibinfo{author}{\bibfnamefont{D.~E.} \bibnamefont{Kharzeev}},
  \bibinfo{author}{\bibfnamefont{L.~D.} \bibnamefont{McLerran}},
  \bibnamefont{and} \bibinfo{author}{\bibfnamefont{H.~J.}
  \bibnamefont{Warringa}}, \bibinfo{journal}{Nucl.Phys.}
  \textbf{\bibinfo{volume}{A803}}, \bibinfo{pages}{227} (\bibinfo{year}{2008}),
  \eprint{0711.0950}.

\bibitem[{\citenamefont{Arsene et~al.}(2005)}]{Arsene:2004fa}
\bibinfo{author}{\bibfnamefont{I.}~\bibnamefont{Arsene}} \bibnamefont{et~al.}
  (\bibinfo{collaboration}{BRAHMS Collaboration}),
  \bibinfo{journal}{Nucl.Phys.} \textbf{\bibinfo{volume}{A757}},
  \bibinfo{pages}{1} (\bibinfo{year}{2005}), \eprint{nucl-ex/0410020}.

\bibitem[{\citenamefont{Back et~al.}(2005)}]{Back:2004je}
\bibinfo{author}{\bibfnamefont{B.}~\bibnamefont{Back}} \bibnamefont{et~al.}
  (\bibinfo{collaboration}{PHOBOS Collaboration}),
  \bibinfo{journal}{Nucl.Phys.} \textbf{\bibinfo{volume}{A757}},
  \bibinfo{pages}{28} (\bibinfo{year}{2005}), \eprint{nucl-ex/0410022}.

\bibitem[{\citenamefont{Adams et~al.}(2005)}]{Adams:2005dq}
\bibinfo{author}{\bibfnamefont{J.}~\bibnamefont{Adams}} \bibnamefont{et~al.}
  (\bibinfo{collaboration}{STAR Collaboration}), \bibinfo{journal}{Nucl.Phys.}
  \textbf{\bibinfo{volume}{A757}}, \bibinfo{pages}{102} (\bibinfo{year}{2005}),
  \eprint{nucl-ex/0501009}.

\bibitem[{\citenamefont{Adcox et~al.}(2005)}]{Adcox:2004mh}
\bibinfo{author}{\bibfnamefont{K.}~\bibnamefont{Adcox}} \bibnamefont{et~al.}
  (\bibinfo{collaboration}{PHENIX Collaboration}),
  \bibinfo{journal}{Nucl.Phys.} \textbf{\bibinfo{volume}{A757}},
  \bibinfo{pages}{184} (\bibinfo{year}{2005}), \eprint{nucl-ex/0410003}.

\bibitem[{\citenamefont{Muller et~al.}(2012)\citenamefont{Muller, Schukraft,
  and Wyslouch}}]{Muller:2012zq}
\bibinfo{author}{\bibfnamefont{B.}~\bibnamefont{Muller}},
  \bibinfo{author}{\bibfnamefont{J.}~\bibnamefont{Schukraft}},
  \bibnamefont{and} \bibinfo{author}{\bibfnamefont{B.}~\bibnamefont{Wyslouch}},
  \bibinfo{journal}{Ann.Rev.Nucl.Part.Sci.} \textbf{\bibinfo{volume}{62}},
  \bibinfo{pages}{361} (\bibinfo{year}{2012}), \eprint{1202.3233}.

\bibitem[{\citenamefont{Fukushima et~al.}(2008)\citenamefont{Fukushima,
  Kharzeev, and Warringa}}]{Fukushima:2008xe}
\bibinfo{author}{\bibfnamefont{K.}~\bibnamefont{Fukushima}},
  \bibinfo{author}{\bibfnamefont{D.~E.} \bibnamefont{Kharzeev}},
  \bibnamefont{and} \bibinfo{author}{\bibfnamefont{H.~J.}
  \bibnamefont{Warringa}}, \bibinfo{journal}{Phys.Rev.}
  \textbf{\bibinfo{volume}{D78}}, \bibinfo{pages}{074033}
  (\bibinfo{year}{2008}), \eprint{0808.3382}.

\bibitem[{\citenamefont{Abelev et~al.}(2009{\natexlab{a}})}]{Abelev:2009ac}
\bibinfo{author}{\bibfnamefont{B.}~\bibnamefont{Abelev}} \bibnamefont{et~al.}
  (\bibinfo{collaboration}{STAR Collaboration}),
  \bibinfo{journal}{Phys.Rev.Lett.} \textbf{\bibinfo{volume}{103}},
  \bibinfo{pages}{251601} (\bibinfo{year}{2009}{\natexlab{a}}),
  \eprint{0909.1739}.

\bibitem[{\citenamefont{Abelev et~al.}(2010)}]{Abelev:2009ad}
\bibinfo{author}{\bibfnamefont{B.}~\bibnamefont{Abelev}} \bibnamefont{et~al.}
  (\bibinfo{collaboration}{STAR Collaboration}), \bibinfo{journal}{Phys.Rev.}
  \textbf{\bibinfo{volume}{C81}}, \bibinfo{pages}{054908}
  (\bibinfo{year}{2010}), \eprint{0909.1717}.

\bibitem[{\citenamefont{Abelev et~al.}(2013)}]{Abelev:2012pa}
\bibinfo{author}{\bibfnamefont{B.}~\bibnamefont{Abelev}} \bibnamefont{et~al.}
  (\bibinfo{collaboration}{ALICE Collaboration}),
  \bibinfo{journal}{Phys.Rev.Lett.} \textbf{\bibinfo{volume}{110}},
  \bibinfo{pages}{012301} (\bibinfo{year}{2013}), \eprint{1207.0900}.

\bibitem[{\citenamefont{Adamczyk et~al.}(2013)}]{Adamczyk:2013hsi}
\bibinfo{author}{\bibfnamefont{L.}~\bibnamefont{Adamczyk}} \bibnamefont{et~al.}
  (\bibinfo{collaboration}{STAR Collaboration}), \bibinfo{journal}{Phys. Rev.}
  \textbf{\bibinfo{volume}{C88}}, \bibinfo{pages}{064911}
  (\bibinfo{year}{2013}), \eprint{1302.3802}.

\bibitem[{\citenamefont{Adamczyk
  et~al.}(2014{\natexlab{a}})}]{Adamczyk:2014mzf}
\bibinfo{author}{\bibfnamefont{L.}~\bibnamefont{Adamczyk}} \bibnamefont{et~al.}
  (\bibinfo{collaboration}{STAR Collaboration}), \bibinfo{journal}{Phys. Rev.
  Lett.} \textbf{\bibinfo{volume}{113}}, \bibinfo{pages}{052302}
  (\bibinfo{year}{2014}{\natexlab{a}}), \eprint{1404.1433}.

\bibitem[{\citenamefont{Adamczyk
  et~al.}(2014{\natexlab{b}})}]{Adamczyk:2013kcb}
\bibinfo{author}{\bibfnamefont{L.}~\bibnamefont{Adamczyk}} \bibnamefont{et~al.}
  (\bibinfo{collaboration}{STAR Collaboration}), \bibinfo{journal}{Phys. Rev.}
  \textbf{\bibinfo{volume}{C89}}, \bibinfo{pages}{044908}
  (\bibinfo{year}{2014}{\natexlab{b}}), \eprint{1303.0901}.

\bibitem[{\citenamefont{Khachatryan et~al.}(2017)}]{Khachatryan:2016got}
\bibinfo{author}{\bibfnamefont{V.}~\bibnamefont{Khachatryan}}
  \bibnamefont{et~al.} (\bibinfo{collaboration}{CMS Collaboration}),
  \bibinfo{journal}{Phys. Rev. Lett.} \textbf{\bibinfo{volume}{118}},
  \bibinfo{pages}{122301} (\bibinfo{year}{2017}), \eprint{1610.00263}.

\bibitem[{\citenamefont{Sirunyan et~al.}(2018)}]{Sirunyan:2017quh}
\bibinfo{author}{\bibfnamefont{A.~M.} \bibnamefont{Sirunyan}}
  \bibnamefont{et~al.} (\bibinfo{collaboration}{CMS}), \bibinfo{journal}{Phys.
  Rev.} \textbf{\bibinfo{volume}{C97}}, \bibinfo{pages}{044912}
  (\bibinfo{year}{2018}), \eprint{1708.01602}.

\bibitem[{\citenamefont{Acharya et~al.}(2018)}]{Acharya:2017fau}
\bibinfo{author}{\bibfnamefont{S.}~\bibnamefont{Acharya}} \bibnamefont{et~al.}
  (\bibinfo{collaboration}{ALICE}), \bibinfo{journal}{Phys. Lett.}
  \textbf{\bibinfo{volume}{B777}}, \bibinfo{pages}{151} (\bibinfo{year}{2018}),
  \eprint{1709.04723}.

\bibitem[{\citenamefont{Kharzeev et~al.}(2016)\citenamefont{Kharzeev, Liao,
  Voloshin, and Wang}}]{Kharzeev:2015znc}
\bibinfo{author}{\bibfnamefont{D.~E.} \bibnamefont{Kharzeev}},
  \bibinfo{author}{\bibfnamefont{J.}~\bibnamefont{Liao}},
  \bibinfo{author}{\bibfnamefont{S.~A.} \bibnamefont{Voloshin}},
  \bibnamefont{and} \bibinfo{author}{\bibfnamefont{G.}~\bibnamefont{Wang}},
  \bibinfo{journal}{Prog. Part. Nucl. Phys.} \textbf{\bibinfo{volume}{88}},
  \bibinfo{pages}{1} (\bibinfo{year}{2016}), \eprint{1511.04050}.

\bibitem[{\citenamefont{Li et~al.}(2016{\natexlab{a}})\citenamefont{Li,
  Kharzeev, Zhang, Huang, Pletikosic, Fedorov, Zhong, Schneeloch, Gu, and
  Valla}}]{Li:2014bha}
\bibinfo{author}{\bibfnamefont{Q.}~\bibnamefont{Li}},
  \bibinfo{author}{\bibfnamefont{D.~E.} \bibnamefont{Kharzeev}},
  \bibinfo{author}{\bibfnamefont{C.}~\bibnamefont{Zhang}},
  \bibinfo{author}{\bibfnamefont{Y.}~\bibnamefont{Huang}},
  \bibinfo{author}{\bibfnamefont{I.}~\bibnamefont{Pletikosic}},
  \bibinfo{author}{\bibfnamefont{A.~V.} \bibnamefont{Fedorov}},
  \bibinfo{author}{\bibfnamefont{R.~D.} \bibnamefont{Zhong}},
  \bibinfo{author}{\bibfnamefont{J.~A.} \bibnamefont{Schneeloch}},
  \bibinfo{author}{\bibfnamefont{G.~D.} \bibnamefont{Gu}}, \bibnamefont{and}
  \bibinfo{author}{\bibfnamefont{T.}~\bibnamefont{Valla}},
  \bibinfo{journal}{Nature Phys.} \textbf{\bibinfo{volume}{12}},
  \bibinfo{pages}{550} (\bibinfo{year}{2016}{\natexlab{a}}),
  \eprint{1412.6543}.

\bibitem[{\citenamefont{Lv et~al.}(2015)}]{Lv:2015pya}
\bibinfo{author}{\bibfnamefont{B.~Q.} \bibnamefont{Lv}} \bibnamefont{et~al.},
  \bibinfo{journal}{Phys. Rev.} \textbf{\bibinfo{volume}{X5}},
  \bibinfo{pages}{031013} (\bibinfo{year}{2015}), \eprint{1502.04684}.

\bibitem[{\citenamefont{Huang et~al.}(2015)}]{Huang:2015eia}
\bibinfo{author}{\bibfnamefont{X.}~\bibnamefont{Huang}} \bibnamefont{et~al.},
  \bibinfo{journal}{Phys. Rev.} \textbf{\bibinfo{volume}{X5}},
  \bibinfo{pages}{031023} (\bibinfo{year}{2015}), \eprint{1503.01304}.

\bibitem[{\citenamefont{Voloshin}(2004)}]{Voloshin:2004vk}
\bibinfo{author}{\bibfnamefont{S.~A.} \bibnamefont{Voloshin}},
  \bibinfo{journal}{Phys.Rev.} \textbf{\bibinfo{volume}{C70}},
  \bibinfo{pages}{057901} (\bibinfo{year}{2004}), \eprint{hep-ph/0406311}.

\bibitem[{\citenamefont{Kharzeev}(2006)}]{Kharzeev:2004ey}
\bibinfo{author}{\bibfnamefont{D.}~\bibnamefont{Kharzeev}},
  \bibinfo{journal}{Phys.Lett.} \textbf{\bibinfo{volume}{B633}},
  \bibinfo{pages}{260} (\bibinfo{year}{2006}), \eprint{hep-ph/0406125}.

\bibitem[{\citenamefont{Bzdak and Skokov}(2012)}]{Bzdak:2011yy}
\bibinfo{author}{\bibfnamefont{A.}~\bibnamefont{Bzdak}} \bibnamefont{and}
  \bibinfo{author}{\bibfnamefont{V.}~\bibnamefont{Skokov}},
  \bibinfo{journal}{Phys. Lett.} \textbf{\bibinfo{volume}{B710}},
  \bibinfo{pages}{171} (\bibinfo{year}{2012}), \eprint{1111.1949}.

\bibitem[{\citenamefont{Deng and Huang}(2012)}]{Deng:2012pc}
\bibinfo{author}{\bibfnamefont{W.-T.} \bibnamefont{Deng}} \bibnamefont{and}
  \bibinfo{author}{\bibfnamefont{X.-G.} \bibnamefont{Huang}},
  \bibinfo{journal}{Phys. Rev.} \textbf{\bibinfo{volume}{C85}},
  \bibinfo{pages}{044907} (\bibinfo{year}{2012}), \eprint{1201.5108}.

\bibitem[{\citenamefont{Bloczynski et~al.}(2013)\citenamefont{Bloczynski,
  Huang, Zhang, and Liao}}]{Bloczynski:2012en}
\bibinfo{author}{\bibfnamefont{J.}~\bibnamefont{Bloczynski}},
  \bibinfo{author}{\bibfnamefont{X.-G.} \bibnamefont{Huang}},
  \bibinfo{author}{\bibfnamefont{X.}~\bibnamefont{Zhang}}, \bibnamefont{and}
  \bibinfo{author}{\bibfnamefont{J.}~\bibnamefont{Liao}},
  \bibinfo{journal}{Phys.Lett.} \textbf{\bibinfo{volume}{B718}},
  \bibinfo{pages}{1529} (\bibinfo{year}{2013}), \eprint{1209.6594}.

\bibitem[{\citenamefont{Wang}(2010)}]{Wang:2009kd}
\bibinfo{author}{\bibfnamefont{F.}~\bibnamefont{Wang}},
  \bibinfo{journal}{Phys.Rev.} \textbf{\bibinfo{volume}{C81}},
  \bibinfo{pages}{064902} (\bibinfo{year}{2010}), \eprint{0911.1482}.

\bibitem[{\citenamefont{Bzdak et~al.}(2010)\citenamefont{Bzdak, Koch, and
  Liao}}]{Bzdak:2009fc}
\bibinfo{author}{\bibfnamefont{A.}~\bibnamefont{Bzdak}},
  \bibinfo{author}{\bibfnamefont{V.}~\bibnamefont{Koch}}, \bibnamefont{and}
  \bibinfo{author}{\bibfnamefont{J.}~\bibnamefont{Liao}},
  \bibinfo{journal}{Phys.Rev.} \textbf{\bibinfo{volume}{C81}},
  \bibinfo{pages}{031901} (\bibinfo{year}{2010}), \eprint{0912.5050}.

\bibitem[{\citenamefont{Schlichting and Pratt}(2011)}]{Schlichting:2010qia}
\bibinfo{author}{\bibfnamefont{S.}~\bibnamefont{Schlichting}} \bibnamefont{and}
  \bibinfo{author}{\bibfnamefont{S.}~\bibnamefont{Pratt}},
  \bibinfo{journal}{Phys.Rev.} \textbf{\bibinfo{volume}{C83}},
  \bibinfo{pages}{014913} (\bibinfo{year}{2011}), \eprint{1009.4283}.

\bibitem[{\citenamefont{Wang and Zhao}(2017)}]{Wang:2016iov}
\bibinfo{author}{\bibfnamefont{F.}~\bibnamefont{Wang}} \bibnamefont{and}
  \bibinfo{author}{\bibfnamefont{J.}~\bibnamefont{Zhao}},
  \bibinfo{journal}{Phys. Rev.} \textbf{\bibinfo{volume}{C95}},
  \bibinfo{pages}{051901} (\bibinfo{year}{2017}), \eprint{1608.06610}.

\bibitem[{\citenamefont{Ajitanand et~al.}(2011)\citenamefont{Ajitanand, Lacey,
  Taranenko, and Alexander}}]{Ajitanand:2010rc}
\bibinfo{author}{\bibfnamefont{N.}~\bibnamefont{Ajitanand}},
  \bibinfo{author}{\bibfnamefont{R.~A.} \bibnamefont{Lacey}},
  \bibinfo{author}{\bibfnamefont{A.}~\bibnamefont{Taranenko}},
  \bibnamefont{and}
  \bibinfo{author}{\bibfnamefont{J.}~\bibnamefont{Alexander}},
  \bibinfo{journal}{Phys.Rev.} \textbf{\bibinfo{volume}{C83}},
  \bibinfo{pages}{011901} (\bibinfo{year}{2011}), \eprint{1009.5624}.

\bibitem[{\citenamefont{Bzdak}(2012)}]{Bzdak:2011np}
\bibinfo{author}{\bibfnamefont{A.}~\bibnamefont{Bzdak}},
  \bibinfo{journal}{Phys.Rev.} \textbf{\bibinfo{volume}{C85}},
  \bibinfo{pages}{044919} (\bibinfo{year}{2012}), \eprint{1112.4066}.

\bibitem[{\citenamefont{Zhao et~al.}(2017)\citenamefont{Zhao, Li, and
  Wang}}]{Zhao:2017nfq}
\bibinfo{author}{\bibfnamefont{J.}~\bibnamefont{Zhao}},
  \bibinfo{author}{\bibfnamefont{H.}~\bibnamefont{Li}}, \bibnamefont{and}
  \bibinfo{author}{\bibfnamefont{F.}~\bibnamefont{Wang}}
  (\bibinfo{year}{2017}), \eprint{1705.05410}.

\bibitem[{\citenamefont{Voloshin}(2010)}]{Voloshin:2010ut}
\bibinfo{author}{\bibfnamefont{S.~A.} \bibnamefont{Voloshin}},
  \bibinfo{journal}{Phys.Rev.Lett.} \textbf{\bibinfo{volume}{105}},
  \bibinfo{pages}{172301} (\bibinfo{year}{2010}), \eprint{1006.1020}.

\bibitem[{\citenamefont{Deng et~al.}(2016)\citenamefont{Deng, Huang, Ma, and
  Wang}}]{Deng:2016knn}
\bibinfo{author}{\bibfnamefont{W.-T.} \bibnamefont{Deng}},
  \bibinfo{author}{\bibfnamefont{X.-G.} \bibnamefont{Huang}},
  \bibinfo{author}{\bibfnamefont{G.-L.} \bibnamefont{Ma}}, \bibnamefont{and}
  \bibinfo{author}{\bibfnamefont{G.}~\bibnamefont{Wang}},
  \bibinfo{journal}{Phys. Rev.} \textbf{\bibinfo{volume}{C94}},
  \bibinfo{pages}{041901} (\bibinfo{year}{2016}), \eprint{1607.04697}.

\bibitem[{\citenamefont{Raman et~al.}(2001)\citenamefont{Raman, Nestor, and
  Tikkanen}}]{Raman:1201zz}
\bibinfo{author}{\bibfnamefont{S.}~\bibnamefont{Raman}},
  \bibinfo{author}{\bibfnamefont{C.~W.~G.} \bibnamefont{Nestor},
  \bibfnamefont{Jr}}, \bibnamefont{and}
  \bibinfo{author}{\bibfnamefont{P.}~\bibnamefont{Tikkanen}},
  \bibinfo{journal}{Atom. Data Nucl. Data Tabl.} \textbf{\bibinfo{volume}{78}},
  \bibinfo{pages}{1} (\bibinfo{year}{2001}).

\bibitem[{\citenamefont{Pritychenko et~al.}(2016)\citenamefont{Pritychenko,
  Birch, Singh, and Horoi}}]{Pritychenko:2013gwa}
\bibinfo{author}{\bibfnamefont{B.}~\bibnamefont{Pritychenko}},
  \bibinfo{author}{\bibfnamefont{M.}~\bibnamefont{Birch}},
  \bibinfo{author}{\bibfnamefont{B.}~\bibnamefont{Singh}}, \bibnamefont{and}
  \bibinfo{author}{\bibfnamefont{M.}~\bibnamefont{Horoi}},
  \bibinfo{journal}{Atom. Data Nucl. Data Tabl.}
  \textbf{\bibinfo{volume}{107}}, \bibinfo{pages}{1} (\bibinfo{year}{2016}),
  \eprint{1312.5975}.

\bibitem[{\citenamefont{Moller et~al.}(1995)\citenamefont{Moller, Nix, Myers,
  and Swiatecki}}]{Moller:1993ed}
\bibinfo{author}{\bibfnamefont{P.}~\bibnamefont{Moller}},
  \bibinfo{author}{\bibfnamefont{J.~R.} \bibnamefont{Nix}},
  \bibinfo{author}{\bibfnamefont{W.~D.} \bibnamefont{Myers}}, \bibnamefont{and}
  \bibinfo{author}{\bibfnamefont{W.~J.} \bibnamefont{Swiatecki}},
  \bibinfo{journal}{Atom. Data Nucl. Data Tabl.} \textbf{\bibinfo{volume}{59}},
  \bibinfo{pages}{185} (\bibinfo{year}{1995}), \eprint{nucl-th/9308022}.

\bibitem[{\citenamefont{Kumar et~al.}(2015)\citenamefont{Kumar, Singh, and
  Patra}}]{Kumar:2014ypa}
\bibinfo{author}{\bibfnamefont{B.}~\bibnamefont{Kumar}},
  \bibinfo{author}{\bibfnamefont{S.~K.} \bibnamefont{Singh}}, \bibnamefont{and}
  \bibinfo{author}{\bibfnamefont{S.~K.} \bibnamefont{Patra}},
  \bibinfo{journal}{Int. J. Mod. Phys.} \textbf{\bibinfo{volume}{E24}},
  \bibinfo{pages}{1550017} (\bibinfo{year}{2015}), \eprint{1409.4645}.

\bibitem[{\citenamefont{Moller et~al.}(2016)\citenamefont{Moller, Sierk,
  Ichikawa, and Sagawa}}]{Moller:2015fba}
\bibinfo{author}{\bibfnamefont{P.}~\bibnamefont{Moller}},
  \bibinfo{author}{\bibfnamefont{A.~J.} \bibnamefont{Sierk}},
  \bibinfo{author}{\bibfnamefont{T.}~\bibnamefont{Ichikawa}}, \bibnamefont{and}
  \bibinfo{author}{\bibfnamefont{H.}~\bibnamefont{Sagawa}},
  \bibinfo{journal}{Atom. Data Nucl. Data Tabl.}
  \textbf{\bibinfo{volume}{109}}, \bibinfo{pages}{1} (\bibinfo{year}{2016}),
  \eprint{1508.06294}.

\bibitem[{\citenamefont{Wang et~al.}(2016)\citenamefont{Wang, Friar, and
  Hayes}}]{Wang:2016rqh}
\bibinfo{author}{\bibfnamefont{X.~B.} \bibnamefont{Wang}},
  \bibinfo{author}{\bibfnamefont{J.~L.} \bibnamefont{Friar}}, \bibnamefont{and}
  \bibinfo{author}{\bibfnamefont{A.~C.} \bibnamefont{Hayes}},
  \bibinfo{journal}{Phys. Rev.} \textbf{\bibinfo{volume}{C94}},
  \bibinfo{pages}{034314} (\bibinfo{year}{2016}), \eprint{1607.02149}.

\bibitem[{\citenamefont{Bender et~al.}(2003)\citenamefont{Bender, Heenen, and
  Reinhard}}]{Bender:2003jk}
\bibinfo{author}{\bibfnamefont{M.}~\bibnamefont{Bender}},
  \bibinfo{author}{\bibfnamefont{P.-H.} \bibnamefont{Heenen}},
  \bibnamefont{and} \bibinfo{author}{\bibfnamefont{P.-G.}
  \bibnamefont{Reinhard}}, \bibinfo{journal}{Rev. Mod. Phys.}
  \textbf{\bibinfo{volume}{75}}, \bibinfo{pages}{121} (\bibinfo{year}{2003}).

\bibitem[{\citenamefont{Erler et~al.}(2012)\citenamefont{Erler, Birge,
  Kortelainen, Nazarewicz, Olsen, Perhac, and Stoitsov}}]{Erler:2012xxx}
\bibinfo{author}{\bibfnamefont{J.}~\bibnamefont{Erler}},
  \bibinfo{author}{\bibfnamefont{N.}~\bibnamefont{Birge}},
  \bibinfo{author}{\bibfnamefont{M.}~\bibnamefont{Kortelainen}},
  \bibinfo{author}{\bibfnamefont{W.}~\bibnamefont{Nazarewicz}},
  \bibinfo{author}{\bibfnamefont{E.}~\bibnamefont{Olsen}},
  \bibinfo{author}{\bibfnamefont{A.~M.} \bibnamefont{Perhac}},
  \bibnamefont{and} \bibinfo{author}{\bibfnamefont{M.}~\bibnamefont{Stoitsov}},
  \bibinfo{journal}{Nature} \textbf{\bibinfo{volume}{486}},
  \bibinfo{pages}{509} (\bibinfo{year}{2012}).

\bibitem[{\citenamefont{Hagen et~al.}(2015)}]{Hagen:2015yea}
\bibinfo{author}{\bibfnamefont{G.}~\bibnamefont{Hagen}} \bibnamefont{et~al.},
  \bibinfo{journal}{Nature Phys.} \textbf{\bibinfo{volume}{12}},
  \bibinfo{pages}{186} (\bibinfo{year}{2015}), \eprint{1509.07169}.

\bibitem[{\citenamefont{Garcia~Ruiz et~al.}(2016)}]{Ruiz:2016gne}
\bibinfo{author}{\bibfnamefont{R.~F.} \bibnamefont{Garcia~Ruiz}}
  \bibnamefont{et~al.}, \bibinfo{journal}{Nature Phys.}
  \textbf{\bibinfo{volume}{12}}, \bibinfo{pages}{594} (\bibinfo{year}{2016}),
  \eprint{1602.07906}.

\bibitem[{\citenamefont{Kortelainen}(2015)}]{Kortelainen:2014uma}
\bibinfo{author}{\bibfnamefont{M.}~\bibnamefont{Kortelainen}},
  \bibinfo{journal}{J. Phys.} \textbf{\bibinfo{volume}{G42}},
  \bibinfo{pages}{034021} (\bibinfo{year}{2015}), \eprint{1409.1413}.

\bibitem[{\citenamefont{Chabanat et~al.}(1998)\citenamefont{Chabanat, Bonche,
  Haensel, Meyer, and Schaeffer}}]{Chabanat:1997un}
\bibinfo{author}{\bibfnamefont{E.}~\bibnamefont{Chabanat}},
  \bibinfo{author}{\bibfnamefont{P.}~\bibnamefont{Bonche}},
  \bibinfo{author}{\bibfnamefont{P.}~\bibnamefont{Haensel}},
  \bibinfo{author}{\bibfnamefont{J.}~\bibnamefont{Meyer}}, \bibnamefont{and}
  \bibinfo{author}{\bibfnamefont{R.}~\bibnamefont{Schaeffer}},
  \bibinfo{journal}{Nucl. Phys.} \textbf{\bibinfo{volume}{A635}},
  \bibinfo{pages}{231} (\bibinfo{year}{1998}), \bibinfo{note}{[Erratum: Nucl.
  Phys.A643,441(1998)]}.

\bibitem[{\citenamefont{Dreizler and Gross}(1990)}]{Dreizler1990nuclear}
\bibinfo{author}{\bibfnamefont{R.~M.} \bibnamefont{Dreizler}} \bibnamefont{and}
  \bibinfo{author}{\bibfnamefont{E.~K.~U.} \bibnamefont{Gross}},
  \emph{\bibinfo{title}{Density Functional Theory: An Approach to the Quantum
  Many-Body Problem}} (\bibinfo{publisher}{Springer, Berlin},
  \bibinfo{year}{1990}).

\bibitem[{\citenamefont{Ring and Schuck}(2000)}]{ring2000nuclear}
\bibinfo{author}{\bibfnamefont{P.}~\bibnamefont{Ring}} \bibnamefont{and}
  \bibinfo{author}{\bibfnamefont{P.}~\bibnamefont{Schuck}},
  \emph{\bibinfo{title}{The Nuclear Many-body Problem}}, Texts and monographs
  in physics (\bibinfo{publisher}{Springer}, \bibinfo{year}{2000}),
  \urlprefix\url{https://books.google.com/books?id=QmQ4nQEACAAJ}.

\bibitem[{\citenamefont{Bartel et~al.}(1982)\citenamefont{Bartel, Quentin,
  Brack, Guet, and Hakansson}}]{Bartel:1982ed}
\bibinfo{author}{\bibfnamefont{J.}~\bibnamefont{Bartel}},
  \bibinfo{author}{\bibfnamefont{P.}~\bibnamefont{Quentin}},
  \bibinfo{author}{\bibfnamefont{M.}~\bibnamefont{Brack}},
  \bibinfo{author}{\bibfnamefont{C.}~\bibnamefont{Guet}}, \bibnamefont{and}
  \bibinfo{author}{\bibfnamefont{H.~B.} \bibnamefont{Hakansson}},
  \bibinfo{journal}{Nucl. Phys.} \textbf{\bibinfo{volume}{A386}},
  \bibinfo{pages}{79} (\bibinfo{year}{1982}).

\bibitem[{\citenamefont{Alver et~al.}(2007)}]{Alver:2006wh}
\bibinfo{author}{\bibfnamefont{B.}~\bibnamefont{Alver}} \bibnamefont{et~al.}
  (\bibinfo{collaboration}{PHOBOS Collaboration}),
  \bibinfo{journal}{Phys.Rev.Lett.} \textbf{\bibinfo{volume}{98}},
  \bibinfo{pages}{242302} (\bibinfo{year}{2007}), \eprint{nucl-ex/0610037}.

\bibitem[{\citenamefont{Miller et~al.}(2007)\citenamefont{Miller, Reygers,
  Sanders, and Steinberg}}]{Miller:2007ri}
\bibinfo{author}{\bibfnamefont{M.~L.} \bibnamefont{Miller}},
  \bibinfo{author}{\bibfnamefont{K.}~\bibnamefont{Reygers}},
  \bibinfo{author}{\bibfnamefont{S.~J.} \bibnamefont{Sanders}},
  \bibnamefont{and}
  \bibinfo{author}{\bibfnamefont{P.}~\bibnamefont{Steinberg}},
  \bibinfo{journal}{Ann.Rev.Nucl.Part.Sci.} \textbf{\bibinfo{volume}{57}},
  \bibinfo{pages}{205} (\bibinfo{year}{2007}), \eprint{nucl-ex/0701025}.

\bibitem[{\citenamefont{Rybczynski and Broniowski}(2011)}]{Rybczynski:2011wv}
\bibinfo{author}{\bibfnamefont{M.}~\bibnamefont{Rybczynski}} \bibnamefont{and}
  \bibinfo{author}{\bibfnamefont{W.}~\bibnamefont{Broniowski}},
  \bibinfo{journal}{Phys. Rev.} \textbf{\bibinfo{volume}{C84}},
  \bibinfo{pages}{064913} (\bibinfo{year}{2011}), \eprint{1110.2609}.

\bibitem[{\citenamefont{Xu et~al.}(2014)\citenamefont{Xu, Pang, and
  Wang}}]{Xu:2014ada}
\bibinfo{author}{\bibfnamefont{H.-j.} \bibnamefont{Xu}},
  \bibinfo{author}{\bibfnamefont{L.}~\bibnamefont{Pang}}, \bibnamefont{and}
  \bibinfo{author}{\bibfnamefont{Q.}~\bibnamefont{Wang}},
  \bibinfo{journal}{Phys. Rev.} \textbf{\bibinfo{volume}{C89}},
  \bibinfo{pages}{064902} (\bibinfo{year}{2014}), \eprint{1404.2663}.

\bibitem[{\citenamefont{Zhu et~al.}(2017)\citenamefont{Zhu, Zhou, Xu, and
  Song}}]{Zhu:2016puf}
\bibinfo{author}{\bibfnamefont{X.}~\bibnamefont{Zhu}},
  \bibinfo{author}{\bibfnamefont{Y.}~\bibnamefont{Zhou}},
  \bibinfo{author}{\bibfnamefont{H.}~\bibnamefont{Xu}}, \bibnamefont{and}
  \bibinfo{author}{\bibfnamefont{H.}~\bibnamefont{Song}},
  \bibinfo{journal}{Phys. Rev.} \textbf{\bibinfo{volume}{C95}},
  \bibinfo{pages}{044902} (\bibinfo{year}{2017}), \eprint{1608.05305}.

\bibitem[{\citenamefont{Abelev et~al.}(2009{\natexlab{b}})}]{Abelev:2008ab}
\bibinfo{author}{\bibfnamefont{B.}~\bibnamefont{Abelev}} \bibnamefont{et~al.}
  (\bibinfo{collaboration}{STAR Collaboration}), \bibinfo{journal}{Phys.Rev.}
  \textbf{\bibinfo{volume}{C79}}, \bibinfo{pages}{034909}
  (\bibinfo{year}{2009}{\natexlab{b}}), \eprint{0808.2041}.

\bibitem[{\citenamefont{Deng and Huang}(2015)}]{Deng:2014uja}
\bibinfo{author}{\bibfnamefont{W.-T.} \bibnamefont{Deng}} \bibnamefont{and}
  \bibinfo{author}{\bibfnamefont{X.-G.} \bibnamefont{Huang}},
  \bibinfo{journal}{Phys. Lett.} \textbf{\bibinfo{volume}{B742}},
  \bibinfo{pages}{296} (\bibinfo{year}{2015}), \eprint{1411.2733}.

\bibitem[{\citenamefont{Lin and Ko}(2002)}]{Lin:2001zk}
\bibinfo{author}{\bibfnamefont{Z.-W.} \bibnamefont{Lin}} \bibnamefont{and}
  \bibinfo{author}{\bibfnamefont{C.}~\bibnamefont{Ko}},
  \bibinfo{journal}{Phys.Rev.} \textbf{\bibinfo{volume}{C65}},
  \bibinfo{pages}{034904} (\bibinfo{year}{2002}), \eprint{nucl-th/0108039}.

\bibitem[{\citenamefont{Lin et~al.}(2005)\citenamefont{Lin, Ko, Li, Zhang, and
  Pal}}]{Lin:2004en}
\bibinfo{author}{\bibfnamefont{Z.-W.} \bibnamefont{Lin}},
  \bibinfo{author}{\bibfnamefont{C.~M.} \bibnamefont{Ko}},
  \bibinfo{author}{\bibfnamefont{B.-A.} \bibnamefont{Li}},
  \bibinfo{author}{\bibfnamefont{B.}~\bibnamefont{Zhang}}, \bibnamefont{and}
  \bibinfo{author}{\bibfnamefont{S.}~\bibnamefont{Pal}},
  \bibinfo{journal}{Phys.Rev.} \textbf{\bibinfo{volume}{C72}},
  \bibinfo{pages}{064901} (\bibinfo{year}{2005}), \eprint{nucl-th/0411110}.

\bibitem[{\citenamefont{Lin}(2014)}]{Lin:2014tya}
\bibinfo{author}{\bibfnamefont{Z.-W.} \bibnamefont{Lin}},
  \bibinfo{journal}{Phys.Rev.} \textbf{\bibinfo{volume}{C90}},
  \bibinfo{pages}{014904} (\bibinfo{year}{2014}), \eprint{1403.6321}.

\bibitem[{\citenamefont{Ma and Lin}(2016)}]{Ma:2016fve}
\bibinfo{author}{\bibfnamefont{G.-L.} \bibnamefont{Ma}} \bibnamefont{and}
  \bibinfo{author}{\bibfnamefont{Z.-W.} \bibnamefont{Lin}},
  \bibinfo{journal}{Phys. Rev.} \textbf{\bibinfo{volume}{C93}},
  \bibinfo{pages}{054911} (\bibinfo{year}{2016}), \eprint{1601.08160}.

\bibitem[{\citenamefont{Wang and Gyulassy}(1991)}]{Wang:1991hta}
\bibinfo{author}{\bibfnamefont{X.-N.} \bibnamefont{Wang}} \bibnamefont{and}
  \bibinfo{author}{\bibfnamefont{M.}~\bibnamefont{Gyulassy}},
  \bibinfo{journal}{Phys.Rev.} \textbf{\bibinfo{volume}{D44}},
  \bibinfo{pages}{3501} (\bibinfo{year}{1991}).

\bibitem[{\citenamefont{Zhang}(1998)}]{Zhang:1997ej}
\bibinfo{author}{\bibfnamefont{B.}~\bibnamefont{Zhang}},
  \bibinfo{journal}{Comput.Phys.Commun.} \textbf{\bibinfo{volume}{109}},
  \bibinfo{pages}{193} (\bibinfo{year}{1998}), \eprint{nucl-th/9709009}.

\bibitem[{\citenamefont{Ma and Zhang}(2011)}]{Ma:2011uma}
\bibinfo{author}{\bibfnamefont{G.-L.} \bibnamefont{Ma}} \bibnamefont{and}
  \bibinfo{author}{\bibfnamefont{B.}~\bibnamefont{Zhang}},
  \bibinfo{journal}{Phys.Lett.} \textbf{\bibinfo{volume}{B700}},
  \bibinfo{pages}{39} (\bibinfo{year}{2011}), \eprint{1101.1701}.

\bibitem[{\citenamefont{Li et~al.}(2016{\natexlab{b}})\citenamefont{Li, He,
  Lin, Molnar, Wang, and Xie}}]{Li:2016flp}
\bibinfo{author}{\bibfnamefont{H.}~\bibnamefont{Li}},
  \bibinfo{author}{\bibfnamefont{L.}~\bibnamefont{He}},
  \bibinfo{author}{\bibfnamefont{Z.-W.} \bibnamefont{Lin}},
  \bibinfo{author}{\bibfnamefont{D.}~\bibnamefont{Molnar}},
  \bibinfo{author}{\bibfnamefont{F.}~\bibnamefont{Wang}}, \bibnamefont{and}
  \bibinfo{author}{\bibfnamefont{W.}~\bibnamefont{Xie}},
  \bibinfo{journal}{Phys. Rev.} \textbf{\bibinfo{volume}{C93}},
  \bibinfo{pages}{051901} (\bibinfo{year}{2016}{\natexlab{b}}),
  \eprint{1601.05390}.

\bibitem[{\citenamefont{Li et~al.}(2017)\citenamefont{Li, He, Lin, Molnar,
  Wang, and Xie}}]{Li:2016ubw}
\bibinfo{author}{\bibfnamefont{H.}~\bibnamefont{Li}},
  \bibinfo{author}{\bibfnamefont{L.}~\bibnamefont{He}},
  \bibinfo{author}{\bibfnamefont{Z.-W.} \bibnamefont{Lin}},
  \bibinfo{author}{\bibfnamefont{D.}~\bibnamefont{Molnar}},
  \bibinfo{author}{\bibfnamefont{F.}~\bibnamefont{Wang}}, \bibnamefont{and}
  \bibinfo{author}{\bibfnamefont{W.}~\bibnamefont{Xie}},
  \bibinfo{journal}{Phys. Rev.} \textbf{\bibinfo{volume}{C96}},
  \bibinfo{pages}{014901} (\bibinfo{year}{2017}), \eprint{1604.07387}.

\bibitem[{\citenamefont{He et~al.}(2016)\citenamefont{He, Edmonds, Lin, Liu,
  Molnar, and Wang}}]{He:2015hfa}
\bibinfo{author}{\bibfnamefont{L.}~\bibnamefont{He}},
  \bibinfo{author}{\bibfnamefont{T.}~\bibnamefont{Edmonds}},
  \bibinfo{author}{\bibfnamefont{Z.-W.} \bibnamefont{Lin}},
  \bibinfo{author}{\bibfnamefont{F.}~\bibnamefont{Liu}},
  \bibinfo{author}{\bibfnamefont{D.}~\bibnamefont{Molnar}}, \bibnamefont{and}
  \bibinfo{author}{\bibfnamefont{F.}~\bibnamefont{Wang}},
  \bibinfo{journal}{Phys. Lett.} \textbf{\bibinfo{volume}{B753}},
  \bibinfo{pages}{506} (\bibinfo{year}{2016}), \eprint{1502.05572}.

\bibitem[{\citenamefont{Poskanzer and Voloshin}(1998)}]{Poskanzer:1998yz}
\bibinfo{author}{\bibfnamefont{A.~M.} \bibnamefont{Poskanzer}}
  \bibnamefont{and} \bibinfo{author}{\bibfnamefont{S.}~\bibnamefont{Voloshin}},
  \bibinfo{journal}{Phys.Rev.} \textbf{\bibinfo{volume}{C58}},
  \bibinfo{pages}{1671} (\bibinfo{year}{1998}), \eprint{nucl-ex/9805001}.

\end{thebibliography}
\end{document}